\documentstyle[11pt,epsfig]{article}
\def\baselinestretch{1.3}
\newcommand{\be}{\begin{equation}}
\newcommand{\ee}{\end{equation}}
\newcommand{\bea}{\begin{eqnarray}}
\newcommand{\eea}{\end{eqnarray}}
\newcommand{\pr}{\partial}

\newcommand{\bse}{\begin{subequations}}
\newcommand{\ese}{\end{subequations}}

\newcommand{\bs}{\begin{slide}}
\newcommand{\es}{\end{slide}}
\def\be {\begin{equation}}
\def\ee {\end{equation}}
\def\ba {\begin{eqnarray}}
\def\ea {\end{eqnarray}}
\def\nn {\nonumber}
\def\bc {\begin{center}}
\def\ec {\end{center}}

\def\p  {\pi}

\def\le {\left}
\def\ri {\right}

\def\f {\frac}

\def\no {\noindent}
\def\bi {\begin{itemize}}
\def\ei {\end{itemize}}

\def\vs {\vspace}

\def\ul {\underline}

\def\bc {\begin{center}}
\def\ec {\end{center}}
\begin{document}

%THE TEXT STARTS HERE
\begin{center}
{\Large\bf Fermion mass splitting, stability and naturalness problems in warped braneworld models}\\ [20mm]
Soumitra SenGupta \footnote{E-mail: tpssg@iacs.res.in} \\
{\em Department of Theoretical Physics,\\
Indian Association for the Cultivation of Science\\
Jadavpur, Kolkata- 700 032, India}
\\[20mm]
\end{center}
\begin{abstract}
We generalize the Randall Sundrum warped braneworld model in six and higher dimension and  propose a resolution
to the mass hierarchy among the standard model fermions. The fine tuning problem in connection with the scalar mass
however is shown to reappear in a new guise in five dimensional warped model when the two form antisymmetric tensor fields, a massless
string mode, propagates in the bulk. Finally the issue of modulus stabilization is re-examined
in presence of a bulk scalar by considering it's back-reaction on the background geometry.
 The role of higher derivative term of the bulk scalar is shown to be crucial to achieve modulus stabilization.   
\end{abstract}
\vskip 1cm
\newpage
\setcounter{footnote}{0}
\def\baselinestretch{1.8}
\section{Background}
Standard model of elementary particles has been extremely 
successful in explaining physical phenomena up to scale close to Tev. 
The hierarchy structure in the background gauge theory 
which refers to the vast disparity between the electroweak scale and the Planck scale 
however gives rise to the the well known fine tuning problem in connection with 
the mass of the only scalar particle in the theory namely Higgs \cite{dreesmartin}.
Higss mass which unlike the fermions is not protected by any symmetry, receives large quadratic quantum correction as,
\ba
\delta m_H^2 &=& \alpha \Lambda^2 \nn
\ea
where $\Lambda$ is the cutoff scale i.e.Planck scale or GUT scale.
To keep $m_H$ within Tev, $\alpha$ must be fine tuned to 
$\sim 10^{-32}$ which is somewhat unnatural.
As a resolution to this so called 'naturalness problem', supersymmetry \cite{dreesmartin} was introduced 
which removes this unnatural feature at the expense of bringing in
the following problems :\\ 
a) It incorporates a large number of superpartners (these extra particles are called sparticles  ) in 
the theory, all of which are so far undetected.\\
b) Particle and the corresponding sparticle must have the same mass.\\  
c) Absence of the superpartners implies that the supersymmetry, even if it is true, is a
broken symmetry at the present energy scale.\\
d) Breakdown of supersymmetry  in turn generates large cosmological constant which is not consistent with it's presently
observed small value.\\
e) The local version of supersymmetry called supergravity \cite{nilles} does allow supersymmetry breaking with zero cosmological constant.
But the resulting theory is not renormalizable unless one embeds the supergravity model in a more 
fundamental theory like string theory \cite{gsw}.\\
Moreover, what will happen if the signature of supersymmetry is not found near Tev scale in the forthcoming experiments? 
One may argue it's presence at a high scale, as required in String theory.
But that will certainly not resolve the hierarchy or the fine tuning problem in connection with the scalar masses.
So unless we subscribe to the exotic ideas like landscape or anthropic principle \cite{anthroland} in favour of
fine tuning or unnaturalness, we will have to look for some alternative paths to resolve this longstanding issue.\\
Theories with extra spatial dimension(s) turn out to be possible alternatives in this direction.
Such theories have 
attracted a lot of attention because of the new geometric approach to solve 
the hierarchy problem. The two most prominent models in this context are 1) ADD model ,proposed by  Arkani-hamed, 
Dimopoulos and Dvali \cite{arkani}, and 
2) RS model, proposed by Randall and Sundrum \cite{lisa}.\\
In ADD model the extra spatial dimension(s) are large and compactified on circles of radii $R$.
The large radius of the extra dimension pulls down the effective higher dimensional
Planck scale i.e the quantum gravity scale.
For two or more extra dimensions, the large radius $R$ can be chosen consistently 
so that the quantum gravity scale $M_d$ at d-dimension and hence the cut-off of the theory $\Lambda$ has the 
desired value $\sim$ Tev.
However in this process it introduces a new hierarchy of length and therefore mass scale in the theory in the form 
of the large radius $R$ ( much larger than the Planck length).\\
In an alternative scenario,
considering one extra spatial dimension Randall and Sundrum proposed a
$5$ dimensional warped geometric model in an anti-deSitter (ADS) bulk spacetime which we describe now.
\section{Randall-Sundrum Model}
In Randall-Sundrum scenario the extra coordinate $y = r \phi$ is compactified on a $S_1/Z_2$ orbifold with
two 3-branes placed at the two orbifold fixed points $\phi = 0,\pi$,
where $r$ is the radius of $S_1$. 
Using $M_{Pl(5)}\equiv M$ the five dimensional action can be written as,
\begin{eqnarray}
S &=& S_{Gravity} + S_{vis} + S_{hid} \nn \\ %+ S_{KR}~ \nn \\
{}\nn \\
\mbox{where,}~~~~S_{Gravity} &=& \int d^4x~r~d{\phi} 
\sqrt{-G}~[ 2M^3R - \underbrace{\Lambda}_{5-dim}]\nn\\
{}\nn\\
S_{vis} &=&  \int d^4x \sqrt{-g_{vis}}~[L_{vis} - V_{vis}] \nn\\
{}\nn\\
S_{hid} &=&  \int d^4x \sqrt{-g_{hid}}~[L_{hid} - V_{hid}] 
\end{eqnarray}
\ul{Metric ansatz:}
\ba
ds^2=e^{-A}~\eta_{\mu\nu}dx^{\mu}dx^{\nu} + 
r^2 d\phi^2
\ea
Warp factor $A(y)$and the brane tensions are found by solving the 5 dimensional
Einstein's equation with orbifolded boundary conditions
\ba
A&=& 2kr\phi \nn\\
{}\nn\\
V_{hid}&=&-V_{vis}= 24 M^3 k 
~~~~~~~~~~~~~~~~~~~~~~
\le[k^2 =\frac{-\Lambda}{24M^3} \ri]    
\ea
The bulk space-time is taken to be anti-desitter  with a negative cosmological constant $\Lambda$.  
\ba
\le( \f{m_H}{m_0}\ri)^2 &=& 
e^{-2A}|_{\phi=\pi} 
= e^{-2kr\p}  \approx (10^{-16})^2 \nn \\
\Rightarrow kr &=& \f{16}{\p}\ln(10)= 11.6279\dots~~\leftarrow \mbox{RS value}
\nn
\ea
with\\
$k \sim M_P$ and $r \sim l_P$\\
So hierarchy problem is resolved without introducing any new scale.

\begin{figure}[h]
\begin{center}
\epsfxsize 2.70 in
\epsfbox{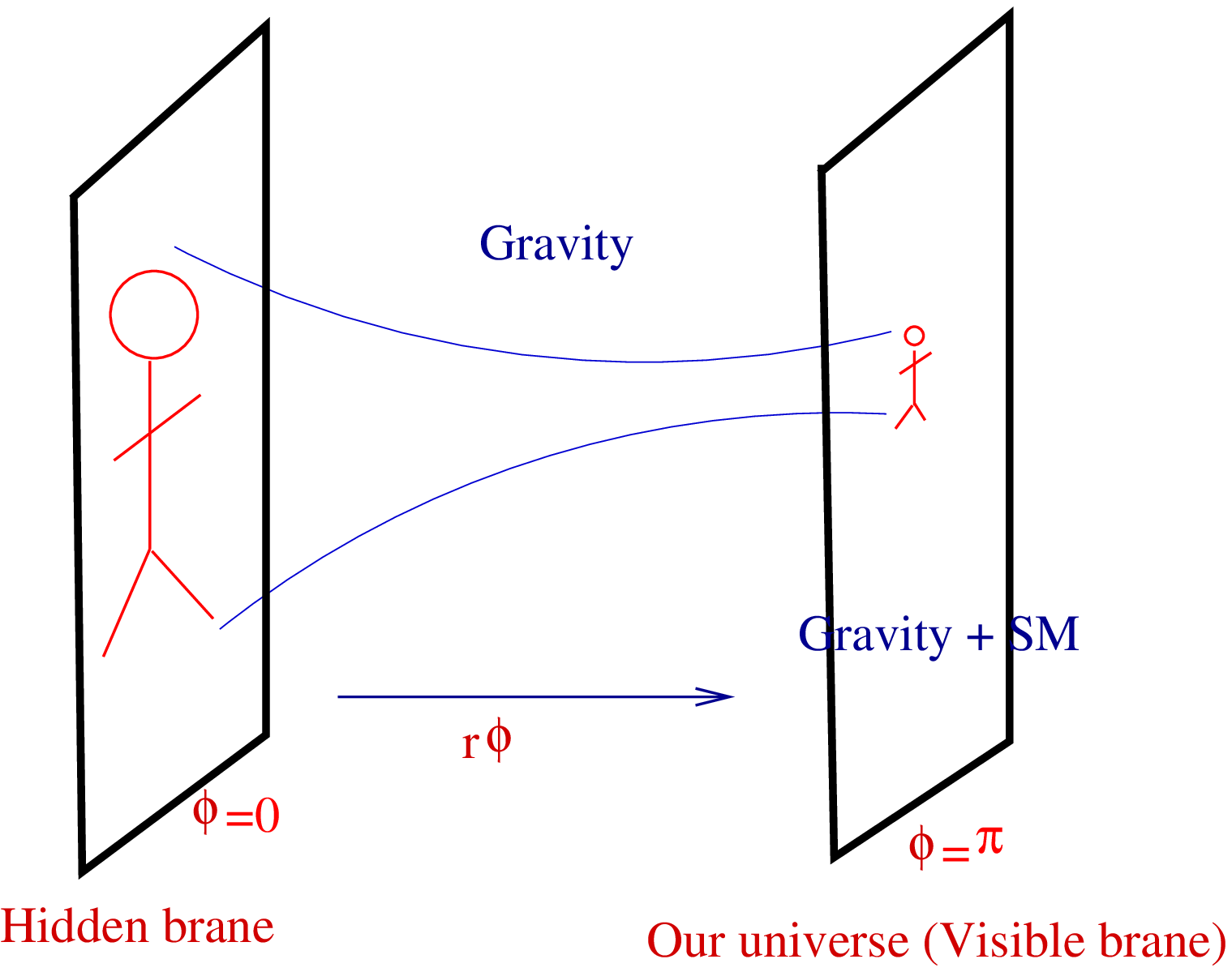}
\end{center}
\end{figure}

A large hierarchy therefore emerges naturally from a small conformal factor. \\
What happens if there are more than one extra warped dimensions \cite{rshigh} ?\\
\section{Generalization to six dimension: Fermion mass splitting}
In a Six dimensional warped space-time
we consider a doubly warped space-time as
$M^{1,5} \rightarrow [M^{1,3} \times S^1/Z_2] \times S^1/Z_2$ \cite{soudebchou}.
The non-compact directions would be denoted by $x^\mu \,
(\mu = 0..3)$ and the orbifolded directions by the angular
coordinates $y$ and $z$ with $R_y$ and $r_z$ as
respective moduli.
Four 4-branes are placed at the orbifold fixed points: \\
$y = 0 ,\pi$ and $z = 0,\pi$ with appropriate brane tensions.
Four 3-branes appear at the four intersection region of these 4-branes.
We also consider an ADS bulk with a negative cosmological constant
$\Lambda$. We thus have a brane-box like space-time as,
\begin{figure}[h]
\begin{center}
\epsfxsize 2.70 in
\epsfbox{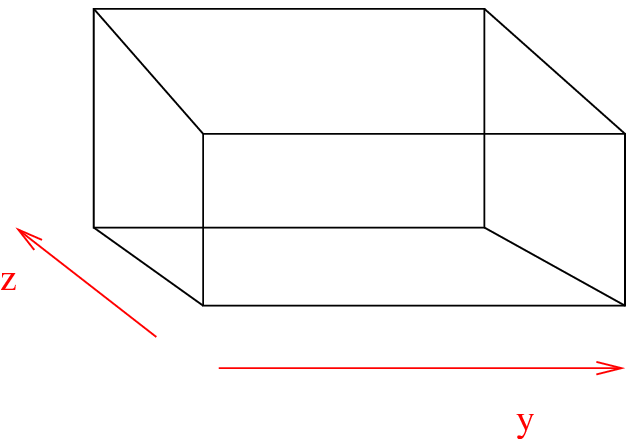}
\end{center}
\end{figure}
\vs{.5cm}
$y$ and $z$ are the compact coordinates

The six dimensional warped metric ansatz:
\ba
ds^2 = b^2(z)[a^2(y)\eta_{\mu\nu}dx^{\mu}dx^{\nu} + R^2_y dy^2] + r^2_z dz^2 \nn
\ea
The total bulk-brane action is thus given by,
\ba
S & = & S_6 + S_5 + S_4 \nn\\
S_6 & = & \int {d^4 x} \, {d y} \, {d z} \, 
          \sqrt{-g_6} \; \left(R_6 - \Lambda \right) \nn\\ 
S_5 & = & \int {d^4 x} \, {d y} \, {d z} \, 
           \left[ V_1 \, \delta(y) + V_2 \, \delta( y - \pi) \right] \nn\\
& + & \int {d^4 x} \, {d y} \, {d z} \, 
           \left[ V_3 \, \delta(z) + V_4 \, \delta(z - \pi) \right] \nn\\
S_4 & = & \int d^4 x \sqrt{-g_{vis}}[{\cal L} - \hat V]  
\ea

In general $V_1, V_2$ are functions of $z$ while $V_3, V_4$ are 
functions of $y$.
The intersecting 4-branes give rise to 3-branes
located at,
$(y, z) = (0,0), (0, \pi), (\pi, 0), (\pi, \pi)$.\\
Substituting the metric in six dimensional Einstein's equation the solutions are:
\ba
a(y) & = & \exp(-c \, y) \nn\\ 
b(z) &=& \frac{\cosh(k \, z)}{\cosh(k \, \pi)} 
\ea
Minimum warping at the 3-brane located at $y = 0, z =\pi$ . Maximum warping at the
3-brane located at $y = \pi, z = 0$. 

Here
\ba
c & \equiv & \frac{R_y \, k}{ r_z \, \cosh(k \, \pi)} \nn\\
k & \equiv &  r_z \, \sqrt{\frac{-\Lambda}{10 \, M^4}}
\ea
Using the orbifolded boundary condition the full
metric thus takes the form,\\
\ba
ds^2 & = & \frac{\cosh^2(k \, z)}{\cosh^2 (k \, \pi)} \,
  \left[ \exp\left(- 2 \, c \, |y| \right)
                \, \eta_{\mu \nu} \, d x^\mu \, d x^\nu  
     + R_y^2 \, d y^2 \right] \nn\\ 
&+& r_z^2 \, d z^2 \ 
\ea
Also the $4+1$ brane tensions become coordinate dependent and are given as,\\
\ba
V_1(z) & = &-V_2(z) = 8M^2 \sqrt{\frac{-\Lambda}{10}} sech(kz) \nn\\ 
V_3(y) & = & 0 \nn \\
V_4(y) & = &-\frac{8 M^4 k}{r_z} \tanh(k \pi)  
\ea
We therefore observe that,
the 3 branes appear at the intersection of the various 4 branes.
There are 4 such 3 branes located at 
$(y, z) = (0,0), (0, \pi), (\pi, 0), (\pi, \pi)$.
The metric on the 3-brane located at $(y = 0, z = \pi)$ suffers no warping.
So it is identified with the Planck brane
Similarly we identify the SM brane with the one at $y= \pi, z = 0$.
Planck scale mass $m_0$ is warped to 
\ba 
m = m_0 \,\frac{r_z \, c}{R_y \, k } \, \exp(-\pi \, c) 
          = m_0 \,  \frac{\exp(-\pi \, c)}{\cosh (k \, \pi)} 
\ea

An interesting picture emerges from this.
To have substantial warping in the $z$-direction (from $z = 0$ to $z = \pi$),
$k \, \pi$ must be substantial, i.e. of same order of magnitude
as in the usual RS case\\
But $c$ is given by 
\ba
c & \equiv & \frac{R_y \, k}{ r_z \, \cosh(k \, \pi)} 
\ea
This immediately means that $c$ must be small for $r_z \sim R_y$\\
Thus we cannot have a large warping in
$y$-direction as well if we want to avoid a new and undesirable
hierarchy between the moduli.
Thus of the two branes located at $y,z =0,0$ and $y,z = \pi ,\pi$,
one must have a natural mass scale close to the Planck scale,
while for the other it is close to the TeV scale.
This resembles to the fine structure splitting of energy levels.

If we repeat this calculation for a seven dimensional
warped space-time , we similarly find, 
eight 3- branes appearing from the intersection of the hyper-surfaces.
Four of these are closed to Tev scale and  
four others are close to Planck scale.
Thus increasing the number of warping in space-time results into
two clusters of branes around Planck and Tev scale.

This leads to phenomenologically interesting consequences like mass splitting of the standard model fermions
on the brane.
The SM-like fields in each of these 3-branes will have apparent mass-scales (on each brane) close to TeV 
with some splitting between them.
To understand this , imagine the SM fermions being defined by 5-dimensional fields 
with $x^\mu$ and $y$ dependence , restricted to the 4-brane say at $z = 0$ which now defines the 
``bulk'' for these fields.
This 4-brane also intersects two other 4-branes at $y = 0$ and $y = \pi$ respectively.
If the major warping has occurred in the $z$-direction, then
the natural mass scale of these fields is still ${\cal O}({\rm TeV})$.
The presence of a $y$-dependence leads to a
non-trivial bulk wave-function.
This,
in turn, changes the overlap of the fermion wave-function with that of
a scalar located on the 3-brane and thus the effective Yukawa
coupling.
The slightly differing interactions on the distant 3-brane
located at two different values of $y$ would result in a hierarchy amongst the effective Yukawa couplings and the fermion masses
on the 3-brane.
Adjusting the parameters suitably the observed mass splitting among the 
different generations of fermions can be explained \cite{fermionmass}. 
\section{Fine tuning problem in presence of higher form bulk field}    
String theory offers a natural explanation for the presence of only the graviton modes in the bulk while all the 
standard model fields on the brane.
Graviton being closed string mode can propagate in the bulk whereas the standard model fields are the open string modes and therefore
their ends are attached to the brane.
However apart from graviton there are other massless closed string excitations also which are free to enter the bulk. 
One of such field is the two form of antisymmetric tensor field namely the Kalb-Ramond (KR) field $B_{MN}$ with the corresponding
third rand antisymmetric tensor field strength $H_{MNL}$ such that $H_{MNL} = \partial_{[M} B_{NL]}$. 
The implications of the presence of such a bulk field in a RS braneworld has already been analyzed in various 
contexts \cite{ssg1}. Here we explore the naturalness issue when such a field exists in the bulk \cite{sauanissg1}\\ 
We begin with the action,
\begin{eqnarray}
\mbox{where,}~~~~S_{Gravity} &=& \int d^4x~r~d{\phi} 
\sqrt{-G}~[ 2M^3R - \underbrace{\Lambda}_{5-d}] \nn\\
{}\nn\\
S_{vis} &=&  \int d^4x \sqrt{-g_{vis}}~[L_{vis} - V_{vis}] \nn\\
{}\nn\\
S_{hid} &=&  \int d^4x \sqrt{-g_{hid}}~[L_{hid} - V_{hid}] \nn \\
{}\nn\\
S_{KR} &=& \int d^4x~r~d{\phi}\sqrt{-G}~[ H_{MNL} H^{MNL}]  
\end{eqnarray}
We also consider the warped metric ansatz:
$ ds^2=e^{-A}~\eta_{\mu\nu}dx^{\mu}dx^{\nu} + 
r^2 d\phi^2~\leftarrow~\mbox{extra dim}~ $

Solving the five dimensional Einstein's equation, the solution for the warp factor turns out to be,
\ba
e^{-A}&=&\frac{\sqrt{b}}{2kr}\cosh{(2kr\phi+2krc)}~ 
\nn\\
{}\nn\\
{}\nn\\
\frac{2kr}{\sqrt{b}}&=&\cosh(2krc)~~~,~~~~\mbox{such that}~~ A(0)=1  \nn\\
{}\nn\\
{}\nn\\
c &=& - \frac{1}{2kr} \tanh^{-1} \left( \frac{V_{hid} }{24 M^3 k} \right)
=  -\pi + \frac{1}{2kr} \tanh^{-1} \left( \frac{V_{vis} }{24 M^3 k} \right)~
\ea
Here $b$ measures the KR Energy density.

The scalar mass warping is now given by,

\ba
 \le(\f{m_H}{m_0}\ri)^2 &=& 
e^{-2A}|_{\phi=\pi} 
=
\frac{\sqrt{b}}{2kr}\cosh\left[2kr\pi + \cosh^{-1} \frac{2kr}{\sqrt{b}}\right]
\nn\\
{}\nn\\
&=&
\left[
\cosh\left(2 kr\pi\right) - \sinh\left( 2kr\pi\right)
\sqrt{1- \frac{b}{(2kr)^2}}
~\right]
\nn\\
{}\nn\\
&\approx& (10^{-16})^2 \nn \\
{}\nn
\ea
Inverting the above expression we obtain,
\ba
b = (2kr)^2
\left[1 -
\left(  
\coth(2kr \pi) - (m_H/m_0)^2 {\rm cosech}(2kr\pi)
\right)^2 
\right]~
\ea
How large $b$ can be to obtain the desired warping from Planck scale to Tev scale? \\

\begin{figure}[h]
\begin{center}
\epsfxsize 2.70 in
\epsfbox{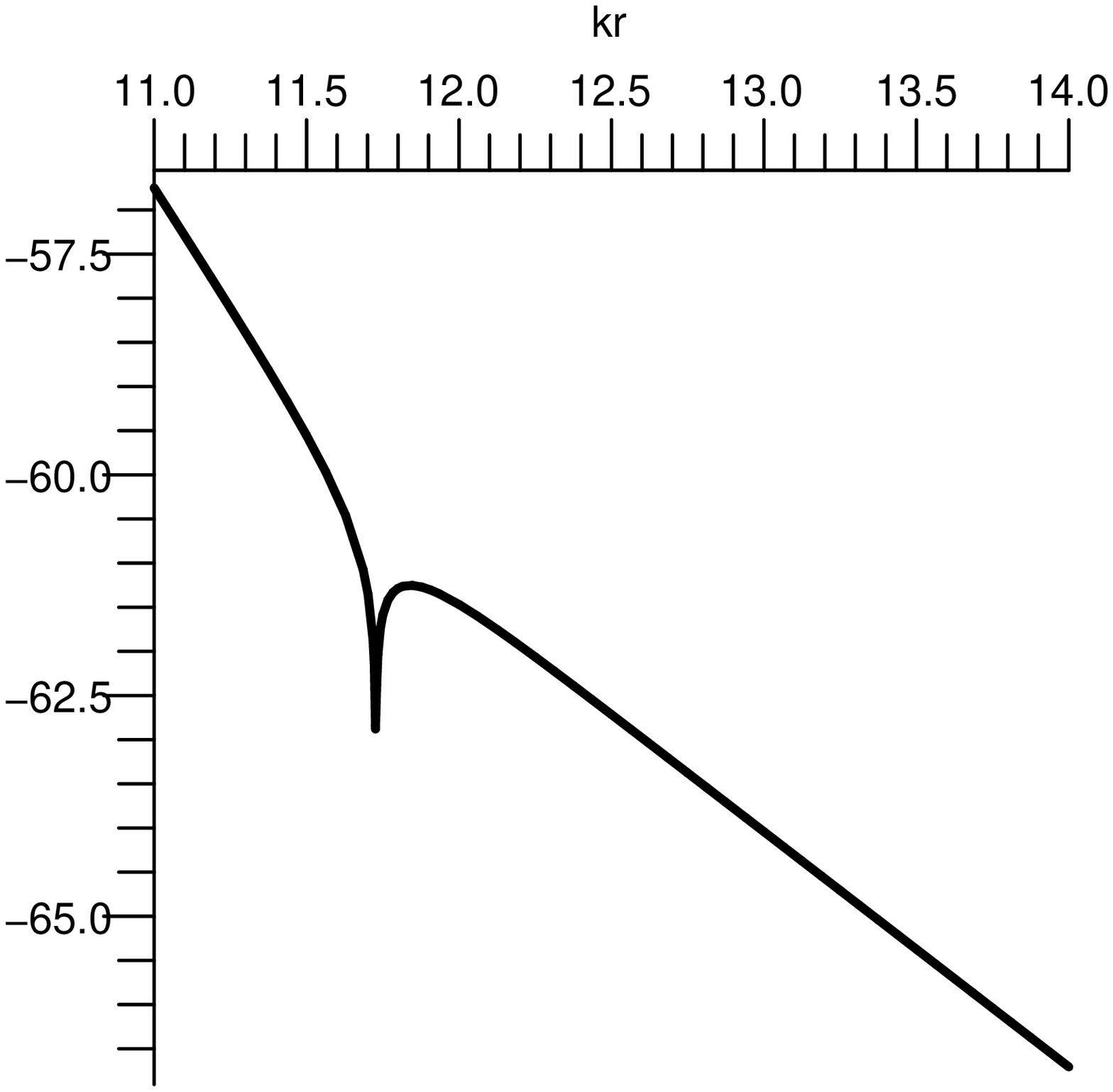}
\end{center}
\end{figure}

\no
$\log|b|$ vs $kr$, for $\f{m_H}{m_0}=10^{-16}$\\
kink $=$ RS value of $kr$ ($b=0$) \\
Left of kink: $kr<$ RS value, $b<0$ \\
Right of kink: $kr>$ RS value, $b>0$

The above figure describes different values for $b$ and the corresponding values for $kr$ which produce the desired warping.
But the maximum possible value for $b$ in this entire range turns out to be,
\ba
b_{max}=10^{-62}
\ea
This indicates extreme fine tuning of $b$ ! 
Thus in order to avoid the fine tuning $\sim 10^{-32}$ of the scalar (Higgs)
mass the warped geometry model was proposed to achieve the desired warping from Planck scale to Tev scale geometrically.
But our result indicates that such geometric warping can be achieved only if the energy density of the bulk
KR field is fine tuned $\sim 10^{-62}$. \\
We have generalized our work by including the dilaton field ( another massless string excitation) also in
the bulk \cite{sauanissg2}. Once again our result revealed that an unnatural fine tuning of the energy density parameter of the
KR field is necessary to achieve the desired warping from the Planck scale to the Tev scale.\\
Thus in the backdrop of a string inspired model the fine tuning problem reappears in a new guise.\\
We now turn our attention to another important issue in the context of braneworld models namely the problem of
modulus stabilization.

\section{Stability problem}
We have already discussed that the desired warping in a RS model can be obtained for  $kr \sim 11$ \\
If $k \sim$ Planck mass, then $r$ should have a stable value near Planck length.
But how to stabilize the modulus $r$ to the desired value ?

A possible resolution to this was first proposed by Goldberg and Wise \cite{gw} as follows :\\
Include a massive scalar field in the bulk of five dimensional gravitational theory
which depends only on the extra coordinate and solve the equation of motion for the scalar field with appropriate 
boundary condition in RS background.
Then Plug in the solution in the action and integrate out the extra coordinate to find an effective
potential for the modulus $r$
and minimize the potential to find the stable value of the modulus $r$.
The crucial assumption in this entire analysis was that the 
'back-reaction of the bulk scalar field on the metric is negligible'.

Subsequently the effect of back-reaction of the scalar field on the 
background metric was found for some specific choices of scalar potential\cite{csakigub}.
However the effective potential for the modulus was not determined to examine the stability
of the braneworld.\\
We now re-examine the stability issue in the back-reacted RS 
model when the scalar field is introduced in the bulk \cite{ssg2}. 

We begin with a very general action :
\ba
S ~=~ \int d^5 x ~\left[ - M^3 R ~+~ F(\phi, X) ~-~ V(\phi)\right] \nn\\ 
 \int d^4 x ~dy~ \sqrt{- g_a} \delta (y - y_a) \lambda_a (\phi)  
\ea
where, $X = \pr_A {\phi} \pr^A {\phi}$\\
`$A$' spans the whole 5-dimensional bulk
space-time. The index `$a$' runs over the brane locations and the corresponding brane 
potentials are denoted by $\lambda_a$. \\
As before the scalar field is assumed to be only the function of extra 
spatial coordinate $y$.

Taking the line element in the form,
\ba
ds^2 ~=~ e^{- 2 A(y)} \eta_{\mu \nu} dx^\mu dx^\nu ~-~ dy^2 ,
\ea
The field equations in the bulk turn out to be
\ba
F_X \phi'' - 2 F_{XX} ~{\phi'}^2 \phi'' &=& 4 F_X \phi' A' - \nn 
\frac {\pr F_X}{\pr \phi}{\phi'}^2 - \nn
\frac 1 2 \left(\frac {\pr F} 
{\pr \phi} - \frac {\pr V}{\pr \phi}\right) \nn\\
{A'}^2 &=& 4 C F_X ~{\phi'}^2 ~+~ 2 C ~\Big[F(X,\phi) ~-~ V(\phi)\Big] \nn\\
A'' &=& 8 C F_X ~{\phi'}^2  ~+~ 4 C \sum_a \lambda_a(\phi) \delta (y - y_a)  
\ea
where $C ~=~ \frac 1 {24 M^3}$

Using the appropriate boundary condition we obtain,\\
\ba 
\frac 1 {\pi} \frac {\pr {V_{eff}}} {\pr r_c} ~=~ -2 e^{-4 A(r_c)} 
\Big[\frac 1 {2 C} ~{A'}^2 ~+~ 4 ~F_{XX}~ {\phi'}^4\Big]_{r_c \pi} 
\ea
So  irrespective 
of the form of the potential, 
stability can be achieved only if the co-efficient 
of the first higher order kinetic term of the bulk scalar field $\phi$
is non-vanishing and is negative.
Therefore canonical scalar field action as considered by 
Goldberger-Wise can not stabilize the modulus and the
higher derivative term plays a crucial role for stability.\\
We consider two examples :\\
1) Along with the higher derivative term ,we take a scalar potential inspired from supergravity models:
\ba
V &=& \frac 1 {16} \left[ \frac 
{\pr W}{\pr \phi} \right]^2 - 2 C W^2 
\ea
with
\ba
W &=& A - B\phi^2 
\ea
A, B are parameters of the superpotential $W$.\\
\vs{.25cm}
2) $F(X,\phi) = - f(\phi) \sqrt{1 - X}$ and $ V(\phi) = 0$\\ which corresponds to a 
tachyon like scalar field in the bulk.\\ 
Our findings are,
\begin{itemize}
\item  Higher derivative terms with appropriate signature is crucial for stability for both
the cases.
\item We can determine the range of stable values of the modulus in both the cases
for different values of the parameters of the potential.
\item It is shown that with appropriate choices of the parameter, the stable value of $r$ can  
produce the desired warping from Planck to Tev scale.
\item We are now trying to generalize this to the higher dimensional warped models. 
\end{itemize}
\section{Summary and Conclusions}
RS model is known to solve the hierarchy problem with gravity in the bulk.
We have shown that a 
multiple warped geometry models ( a generalization of the RS model to higher dimension )can provide a clue to the
fermion mass hierarchy observed in the standard model.
Interesting graviton tower appears in this multiple warped model with possibly interesting
phenomenology.

However in a string based model where two form antisymmetric field is present in the bulk, 
the desired warping can only be achieved by an
extreme fine tuning of KR field as  $1$ part in $10^{62}$. 
Does this signal the {\it Return of the hierarchy/fine-tuning problem in a new guise}?

We then re-examined the issue of modulus stabilization in a back-reacted RS model.
Our results clearly indicate that the stability is not possible by introducing a bulk scalar
with canonical action. Stability and the desired warping however can be found with
higher derivative terms of the scalar field. 
Such a stable value is shown to depend on the coefficient of the 
higher derivative term.\\
In conclusion the RS model perhaps is not the answer to the gauge hierarchy/fine tuning  problem
when considered in string theoretic scenarios.\\
Is supersymmetry therefore so far the best way to resolve the gauge hierarchy/fine tuning problem?

%\end{Large}

%\end{document}

\end{document}